\documentclass[10pt]{article}
\usepackage{amsmath,amsfonts,amssymb,amsthm,bbm}

\usepackage{amssymb,graphicx}

\usepackage{amssymb}

\newcommand{\koniec}{\begin{flushright}  $\Box $ \end{flushright}}

\newcounter{mnotecount}[section]

\renewcommand{\themnotecount}{\thesection.\arabic{mnotecount}}

\newcommand{\mnote}[1]
{\protect{\stepcounter{mnotecount}}$^{\mbox{\footnotesize
$
\bullet$\themnotecount}}$ \marginpar{
\raggedright\tiny\em
$\!\!\!\!\!\!\,\bullet$\themnotecount: #1} }

\newcommand{\CP}{\mathbb{CP}}

\let\a=\alpha 
\let\b=\beta
\def\r{\rho}

\def\ov{\overline}

\usepackage{amsmath,amsfonts,amssymb,amsthm,bbm}

\usepackage{cite}
\usepackage{graphicx}
\usepackage[latin1]{inputenc}
\usepackage{amsthm}
\usepackage{hyperref}
\usepackage{color}
\usepackage{enumerate}

\usepackage{pdfpages}

\newcommand{\Ref}[1]{(\ref{#1})}
\newcommand{\ZZ}{{\cal Z}}
\newcommand{\C}{{\mathbb C}}

\newcommand{\Z}{{\mathbb Z}}
\newcommand{\M}{{\mathbb M}}

\newcommand{\SU}{{SU}}

\newcommand{\SL}{{SL}}

\newcommand{\U}{{U}}

\newcommand{\su}{{\mathfrak{su}}}

\renewcommand{\u}{{\mathfrak u}}

\newcommand{\be}{\begin{equation}}
\newcommand{\ee}{\end{equation}}
\newcommand{\bea}{\begin{eqnarray}}
\newcommand{\eea}{\end{eqnarray}}
\newcommand{\nn}{\nonumber}
\newcommand{\bit}{\begin{itemize}}
\newcommand{\eit}{\end{itemize}}

\newcommand{\tr}{{\rm Tr}}
\newcommand{\f}{\frac}
\newcommand{\tl}{\tilde}

\newcommand{\Id}{\mathbbm{1}}

\newcommand{\mat} [4] {\left ( \begin{array}{cc}{#1}&{#2}\\{#3}&{#4} \end{array} \right ) }

\renewcommand{\a}{\alpha} \renewcommand{\b}{\beta} \newcommand{\g}{\gamma}  
\renewcommand{\d}{\delta}  \newcommand{\eps}{\epsilon}

\let\m=\mu       \let\r=\rho \let\om=\omega
       
\let\G=\Gamma   \let\Th=\Theta

\def\T{\mathbbm T}

\begin{document}
\date{\small{v1: January 23, 2019, v2: April 30, 2019}}
\title{An octahedron of complex null rays, and conformal symmetry breaking}

\author{{Maciej Dunajski$^{a}$, Miklos L$\mathring{\mathrm{a}}$ngvik$^{b}$ and Simone 
Speziale$^{c}$}
\smallskip \\ 
\small{$^{a}$ DAMTP, Centre for Mathematical Sciences, Wilberforce Road, Cambridge CB3 0WA, UK} \\
\small{$^{b}$ Department of Physics, University of Helsinki,
P.O. Box 64, FIN-00014 Helsinki, Finland, and} \\
\small{Ash\"ojdens  grundskola, Sturegatan 6, 00510 Helsingfors, Finland} \\
\small{$^{c}$ Centre de Physique Th\'{e}orique, Aix Marseille Univ., Univ. de Toulon, CNRS, Marseille, France}}

\maketitle

\begin{abstract} 
We show how the manifold $T^*SU(2, 2)$ arises as a symplectic reduction 
from eight copies of the twistor space. Some of the constraints in the 
twistor space correspond to an octahedral configuration
of twelve complex light rays in the Minkowski space. We discuss a mechanism
to break the conformal symmetry down to the twistorial parametrisation
of $T^*SL(2, \C)$ used in loop quantum gravity.
\vskip5pt
\end{abstract}  
\begin{center} 
{\em In memory of Sir Michael Atiyah (1929--2019)}
\end{center}

\section{Introduction}
A twistor space  $\T=\C^4$ is a complex--four dimensional 
vector space equipped with a
pseudo--Hermitian inner product $\Sigma$ of signature $(2, 2)$, 
and the associated natural symplectic structure \cite{Roger1, PenroseRindler2}.
In \cite{Tod76} Tod has shown that the symplectic form
induced on a five--dimensional real surface of projective twistors
which are isotropic with respect to $\Sigma$ 
coincides with the symplectic form on the space of null
geodesics in the $3+1$--dimensional Minkowski space $\M$. It the same
paper it was demonstrated that the Souriau symplectic form \cite{souriau}
on the space of massive particles in $\M$ with spin 
arises as a symplectic reduction from $\T\times \T$.
In a 
{different} context it was shown 
in \cite{twigeo2} and \cite{WielandTwistors, IoWolfgang, IoTwistorNet} that a cotangent bundle $T^*\mathcal{G}$ to a Lie group $\mathcal{G}$
arises from $\T$ if $\mathcal{G}=SU(2)$ and $\T\times \T$ if $\mathcal{G}=SL(2, \C)$. In these references the Darboux coordinates were constructed from spinors and twistors respectively.

The aim of this paper is to extend these constructions to the case when 
$\mathcal{G}=SU(2, 2)$, the covering group of the conformal 
group $SO(4, 2)/\Z_2$ of $\M$. The starting point for our construction will
be the 64--dimensional real vector space consisting of two copies
of ${\T^4}\equiv\T\times \T\times \T\times \T$. The symplectic reduction from 
${\T^4}\times {\T^4}$ to the 30--dimensional manifold 
$T^*SU(2, 2)$ will be realised
by imposing a set of constraints: the second class {\em incidence constraints} stating that
the four twistors in each copy of ${\T^4}$ are non--isotropic and pairwise 
orthogonal with respect to $\Sigma$, and the first class helicity {and} phase constraints (see \S \ref{sec_ss} for details).

All these constraints are conformally invariant when expressed in the Minkowski space $\M$.
The incidence constraints have a natural geometric intepretations in terms of four twistors
in a single twistor space: they describe a  tetrahedron in $\T$, whose 
vertices correspond to twistors and faces to dual
twistors. This configuration is self--dual
in a sense to be made precise in \S\ref{sd_section}.
In the Minkowski space this tetrahedron corresponds to an octahedral configuration
of twelve complex null rays. 

Our main motivation to perform this analysis is to further explore the mathematical relations between twistor theory and loop quantum gravity (LQG)
\cite{WielandTwistors,IoWolfgang,IoTwistorNet,IoMiklos,IoFabio,Wieland:2017zkf,Wieland:2017cmf}. The building blocks of LQG are Penrose's SU(2) spin networks, with an important conceptual difference. Penrose regarded the quantum labels on these networks to describe only the conformal structure of spacetime, specifically angles \cite{spin_net}. To introduce a notion of scale, he envisaged extending the theory to the Poincar\'e group, or better to SU(2,2) that is semi-simple. The associated conformal spin networks and their geometric interpretation have never been used in quantum gravity models, but these ideas then flew into the construction of twistors, which are SU(2, 2) spinors. In LQG on the other hand, the use of Ashtekar-Barbero variables underpinning the theory allows to interpret the SU(2) Casimir directly in terms of areas, thus introducing scales. It is nonetheless still an open and intriguing question to develop Penrose's original program and show if and how a notion of scale relevant for quantum gravity can be introduced via the translation and dilation generators of SU(2,2), and how it can be compared with the one used in LQG through some mechanism for conformal symmetry breaking. To that end, one needs to establish
a precise relation between SU(2,2) spin networks and the SU(2) ones used in LQG. 
As a first step in this direction, we consider the classical counterpart to this question. Recall in fact that the spin network Hilbert space $L^2[\mathcal{G},d\m_{\rm Haar}]$ with its holonomy-flux algebra is, for any Lie group $\mathcal{G}$, the quantization of the canonical Poisson algebras of the cotangent bundle $T^*\mathcal{G}$. We can thus ask how the classical phase space $T^*\SU(2)$ used in LQG can be embedded in $T^*\SU(2, 2)$.  
Our work answers this question.
We provide a uniform parametrization of $T^*\SU(2)$, $T^*\SL(2,\C)$ and $T^*\SU(2, 2)$ in terms of twistors. The embedding is identified by a hypersurface where the dilatation generators match.  
{This matching breaks conformal symmetry in a way that, unlike in standard twistor theory, does not require introducing the infinity twistor.}

The paper is organised as follows. In the next section we shall
introduce the twistor space $\T$ of the Minkowski space, and to 
prepare the ground for the  constraint analysis we shall construct
an octahedral configuration of complex rays in $\M_\C$ out of four non--null
incident twistors. In \S\ref{main_sec} we shall consider
a set of constraints on a product ${\T^4}\times{\T^4}$ of eight twistor spaces, and implement a symplectic reduction to the canonical symplectic form
on $T^*SU(2, 2)$. Finally in \S\ref{conf_sym_sec} we shall comment on the 
conformal symmetry breaking of our construction down to $T^*SL(2, \C)$ and
$T^*SU(2)$, 
and on the physical applications of our results.

\vskip6pt {\it Sir Michael Atiyah died on the 11th of January 2019. 
Sir Michael was a giant of 20th century mathematics, and one of the key
contributors in the development  of twistor theory \cite{AHS}. We dedicate
this paper to his memory.} 

\section{Twelve complex null rays from a twistor tetrahedron}
\label{sd_section}
The twistor programme of Roger Penrose \cite{Roger1}
is a geometric framework for
physics that aims to unify general relativity and quantum mechanics 
with space--time events being derived objects that correspond 
to compact holomorphic curves in a complex manifold known as the projective twistor space $P\T$.
There are now many applications of twistors in pure mathematics, and 
theoretical physics  (see \cite{ADM} for a recent review). Our presentation 
below focuses on the simplest case of twistor space corresponding to the flat 
Minkowski space.

A twistor space  $\T=\C^4$ is a complex four--dimensional 
vector space equipped with a
pseudo--hermitian inner product $\Sigma$ of signature $(2, 2)$
\be
\label{inner_p}
\Sigma(Z, {Z})=Z^1\bar{Z}^3+Z^2\bar{Z}^4+Z^{3}\bar{Z}^1+Z^4\bar{Z}^2,
\ee
where $(Z^1, Z^2, Z^3, Z^4)$ are coordinates in $\T$.

Let $\T^*$ be the dual vector space, and let $P\T=\CP^3$ be a projectivisation of 
$\T$. Let $PN=\{Z\in P\T, \Sigma(Z, {Z})=0\}$ be a real 5-dimensional surface
in $P\T$. The points in $PN$ are referred to as null twistors and correspond 
to real null light rays in Minkowski space \cite{PenroseRindler2, dunajski_book}. 
A dual twistor
$W\in P\T^*$ corresponds to a projective plane  
$W\equiv\{Z\in P\T, W(Z)=0\}\subset P\T$. We say that $Z$ and $W$ are 
incident if  $Z$ lies on the plane given by $W$. In this case the 
$\alpha$-plane $Z$ and the $\beta$-plane $W$ in the complexified Minkowski space
$\M_\C$ intersect in a null geodesic.
\begin{center}
\includegraphics[width=8cm,height=8cm,angle=0]{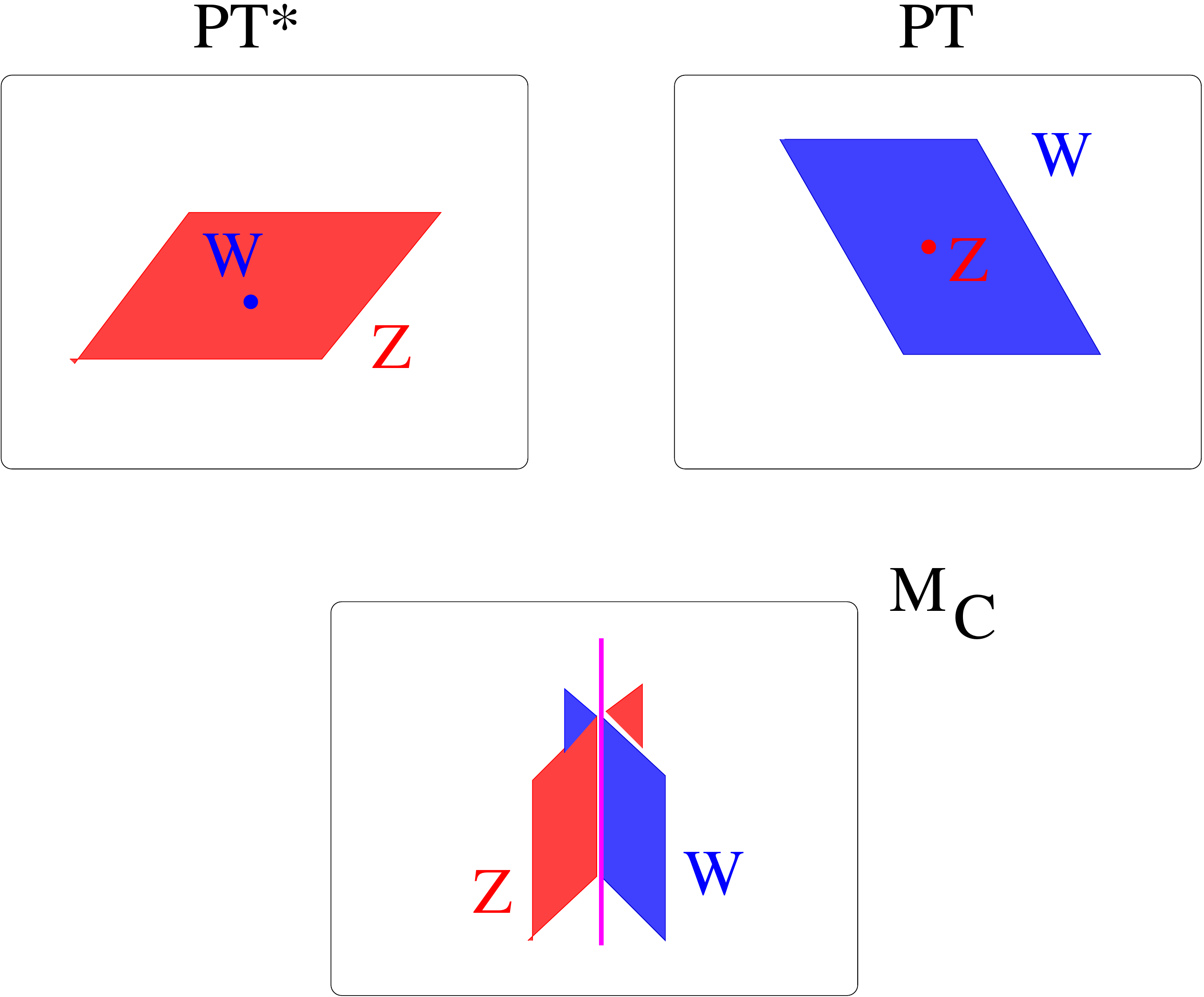}
\begin{center}
{{\bf Fig. 1}. {Twistor incidence and null rays.}}
\end{center}
\end{center}
Let $Z\in P\T$ be a non-null twistor. We can use $\Sigma$ to identify
the conjugation $\ov{Z}$ with an element of $P\T^*$. 
Thus $\ov{Z}$ is a dual twistor corresponding to
a two--plane $\CP^2$ in the projective twistor space $P\T$. The plane
$\ov{Z}$ intersects the hyper-surface $PN$ in a real three--dimensional 
surface -  the Robinson congruence in the Minkowski space.
The point $Z$ lies on the plane $\ov{Z}$ iff $Z\in PN$. Then the  complex $\alpha$--plane
$Z$ meets the compex $\beta$--plane $\ov{Z}$ in a real null geodesics in $\M_\C$. 
Assume that this does not happen.
\vskip5pt
Let $Z_1, Z_2$  be two non-null twistors. They are incident if $Z_1$ belongs
to the plane $\ov{Z}_2$ in $P\T$ and $Z_2$ belongs to the plane $\ov{Z}_1$. The two planes
$\ov{Z}_1$ and $\ov{Z}_2$ intersect in a holomorphic line $\ov{X}_{12}$ in $P\T$.
Now let us add a non--null twistor $Z_3$. 
It will be incident with $Z_1$ and $Z_2$ only if
it lies on the holomorphic line $\ov{X}_{12}$ above. Thus, given an incident non--null
pair $Z_1, Z_2$, there exists a one--parameter family of $Z_3\in P\T$ such
that $Z_1, Z_2, Z_3$ are mutually incident. The plane $\ov{Z}_3$ intersects the line
$\ov{X}_{12}$ in a unique point $Z_4$ and the four twistors $Z_i, i=1, \dots, 4$
satisfy
\be
\label{inci}
\Sigma(Z_i, {Z}_j)=0, \quad i\neq j.
\ee
It is not possible to construct a set $\{Z_i\}$ of more than four twistors
such that (\ref{inci}) holds: the four twistors 
correspond to four vertices
of a tetrahedron in $P\T$. The dual twistors are the faces of this tetrahedron.
A fifth twistor $Z_5$ can not be added in a way that
makes all sets of three points co-linear (or such that the plane $\ov{Z}_5$
intersects all faces of the tetrahedron).
\begin{center}
\includegraphics[width=6cm,height=6cm,angle=0]{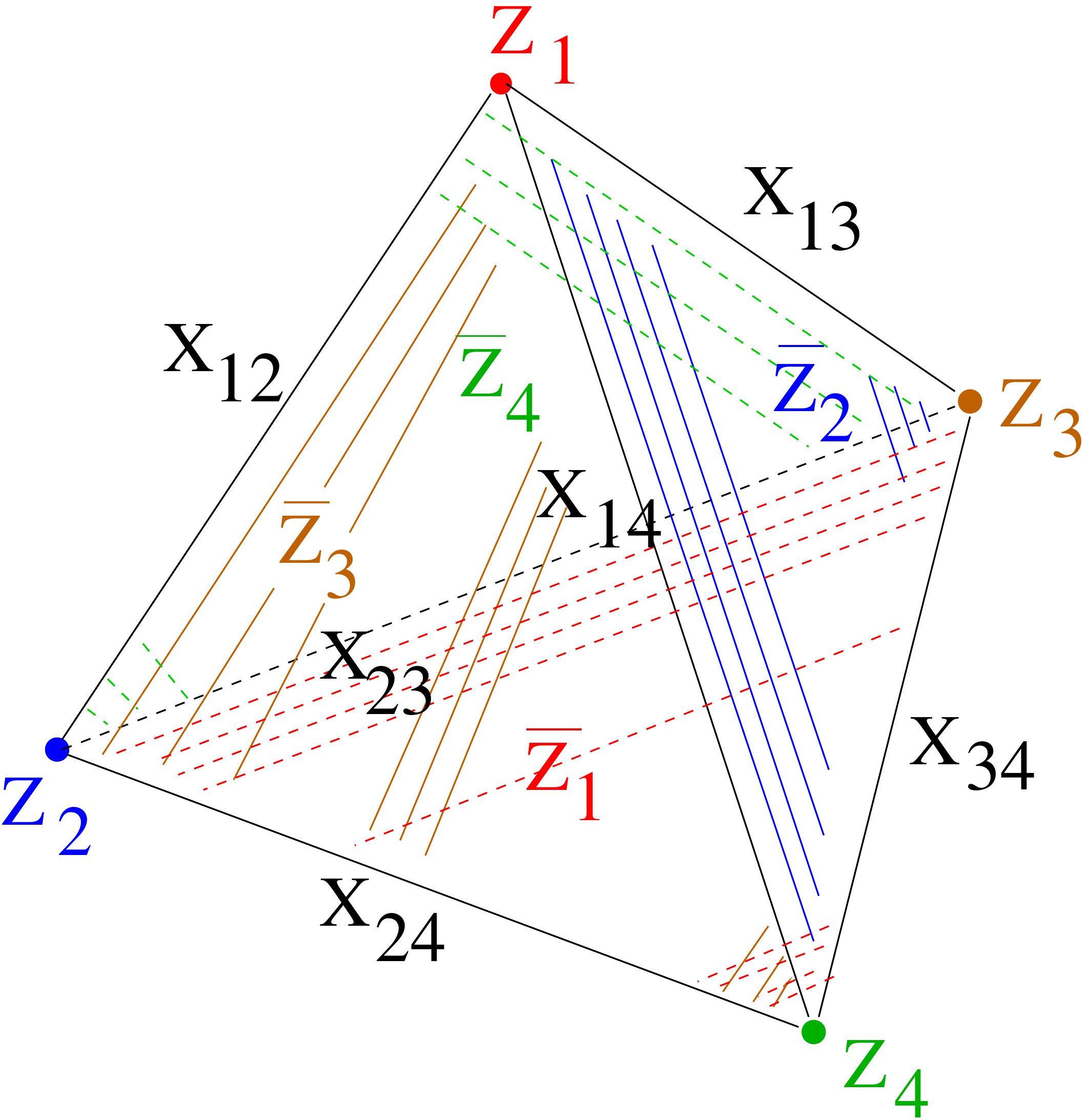}
\begin{center}
{{\bf Fig. 2}. {\em A tetrahedron in $P\T$. Vertices are 
the incident
twistors, faces are the dual twistors and the edges
are lines corresponding to points of intersections of $\alpha$--planes
in $M_\C$. }}
\end{center}
\end{center}
Let $X_{ij}\cong \CP^1 $ be a holomorphic line in $P\T$ joining two twistors
$Z_i$ and $Z_j$, and let  $\ov{X}_{ij}\cong \CP^1$ be a holomprhic line in 
$PT$ arising as the intersection of the planes $\ov{Z}_i$ and $\ov{Z}_j$.
Then
\[
X_{12}=\ov{X}_{34}, \quad X_{13}=\ov{X}_{24}, \quad \mbox{etc},
\]
which resembles the self--duality condition.
The line $X_{ij}$ corresponds to a unique point of intersection
of two $\alpha$--planes $Z_i$ and $Z_j$ in $\M_\C$. Similarly, 
the line $\ov{X}_{ij}$ corresponds to a point of intersection of two 
$\beta$ planes $\ov{Z}_i$ and $\ov{Z}_j$ in $\M_\C$. If $i\neq j$, then the
$\alpha$--plane $Z_i$ intersects the $\beta$--plane $\ov{Z}_j$ in a complex
null geodesics (a light ray) $R_{i\ov{j}}$. 
 \begin{center}
\includegraphics[width=4cm,height=4cm,angle=0]{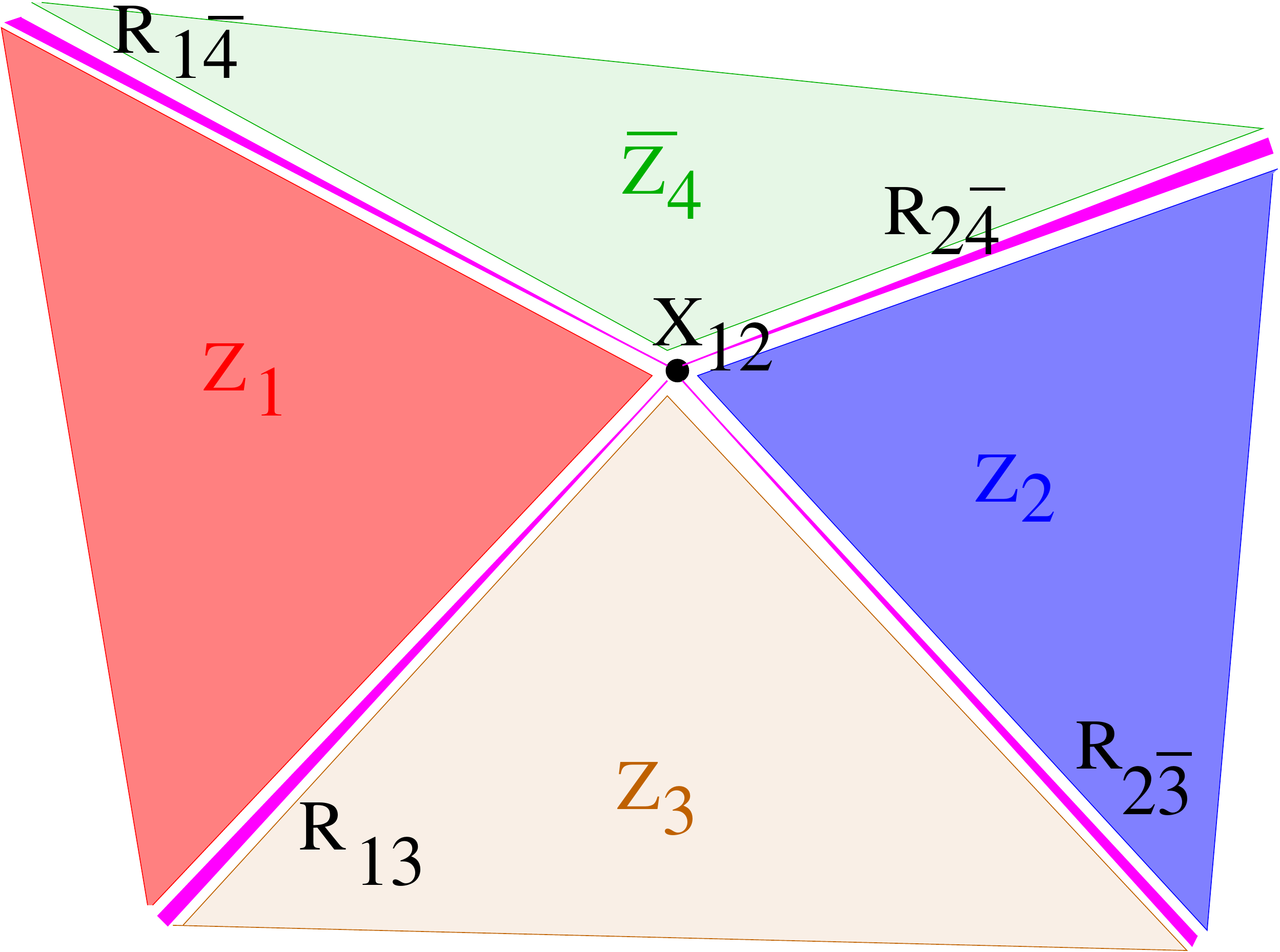}
\begin{center}
{{\bf Fig. 3}. {\em An intersection of an $\alpha$--plane $Z_i$ with a 
$\beta$--plane $\ov{Z}_j$ is a null ray $R_{i\ov{j}}$.}}
\end{center}
\end{center}
This leads to the octahedral 
configuration of twelve complex null rays arising as 
intersections of incident $\alpha$ and $\beta$--planes.  The six
vertices of the resulting octahedron ${\bf O}$ in $\M_\C$ correspond to the six lines
$X_{ij}$ in $P\T$. The twelve edges of ${\bf O}$ are complex 
null rays.
\begin{center}
\includegraphics[width=6cm,height=6cm,angle=0]{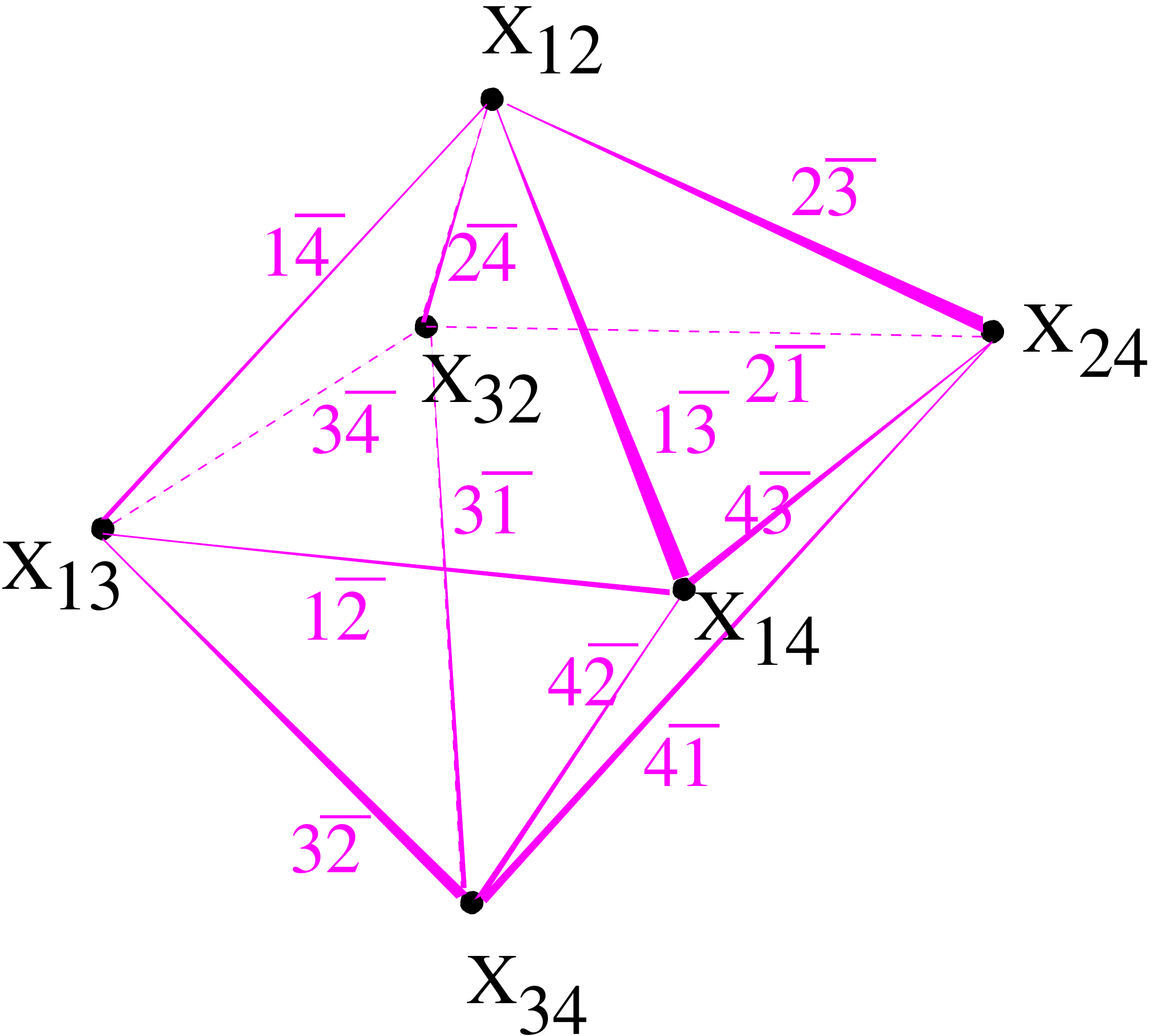}
\begin{center}
{{\bf Fig. 4}. {\em The octahedral configuration of twelve complex null rays
in $\M_\C$.}}
\end{center}
\end{center}
\section{Symplectic reduction to $T^*SU(2, 2)$}
\label{main_sec}
Our strategy will be to pick two linearly independent sets of four
twistors, and construct an element $G$ of $GL(4, \C)$ mapping one set
to the other. We shall then impose a set of constraints on both sets which will guarantee that $G$ is unitary, and has unit determinant. Some of these
constraints will be first class, and some second class with respect
to the twistor symplectic structure, and we will show (by explicit computation
of Poisson brackets) how the symplectic structure on $T^*SU(2, 2)$ arises
as symplectic reduction from the symplectic structure on eight copies of
$\T$.
\subsection{Notation}
In what follows we shall denote components of a twistor $Z\in \T$ by
$Z^{\alpha}, \alpha=1,\dots, 4$, and components of the corresponding 
dual twistor by $\ov{Z}_\alpha\equiv\Sigma_{\alpha\dot\beta}\ov{Z}{}^{\dot\beta}$
where
\be\label{eta}
\Sigma_{\a\dot\b} = \mat{0}{\Id_2}{\Id_2}{0}
\ee
is a matrix of the $(2, 2)$ inner product $\Sigma$ from \S\ref{sd_section}.
The imaginary part of $\Sigma$  gives the twistor space a Poisson structure
\be\label{ZZ}
\{Z^\a, \ov{Z}_{\b} \} = i \d^\a_\b,
\ee 
which  is invariant under $SU(2, 2)$ transformations of $\T$. {These are}
generated via a Hamiltonian action, 
\be
\label{MZ}
\{M^{ab}, Z^\a\} = \G^{ab\a}{}_\b Z^\b, \quad\mbox{where}\quad M^{ab} := \ov{Z}_{\a} \G^{ab\a}{}_\b Z^\beta, \quad
\mbox{and}\quad a, b=0, \dots, 5.
\ee
The matrices $\G^{ab}\equiv (1/2)[\G^a, \G^b]$ are constructed out of the six generators $\G^a$ of the Clifford
algebra in $(4+2)$ dimensions, and they form a representation of $\mathfrak{spin}(4,2)$. They also form
15 out of the 16 generators of $\mathfrak{u}(2, 2)$. The last one is the trivial identity element (normalized by 1/2), and correspond to the helicity,
\be\label{defU}
U:= \f12 \ov{Z}_{\a}\d^{\a}{}_\b Z^\b = s.
\ee
\subsection{Unitary transformations}
Let $(Z_1, Z_2, Z_3, Z_4)\in {\T^4}\equiv \T\times \T\times \T\times \T$
be four twistors such that the {\em holomorphic volume}
\be\label{defZ}
\ZZ:= \f1{4!}\eps^{ijkl}\eps_{\a\b\g\d} Z^\a_i Z^\b_j Z^\g_k Z^\d_l\neq 0.
\ee
Here, for each fixed $i=1, \dots, 4$ the symbol $Z^\a_i$ denotes the four 
components of $Z_i$ with respect to the standard basis of $\T$.
We shall use a summation convention with the Latin indices
$\alpha, \beta, \dots$, and our formulae will be $SU(2, 2)$ invariant
in these indices. The Greek indices $i, j, k, \dots $ are 
reminiscent of the internal twistor indices in the twistor particle programme
\cite{particles1, particles2}.
Parts of our construction will break the internal symmetry, in which case we will write explicitly the sums over internal indices. This makes the resulting formulae somewhat ugly.
We  set 
\be\label{dualbasis}
Z^i_\a := \f1{6\ZZ} \eps^{ijkl}\eps_{\a\b\g\d} Z^\b_j Z^\g_k Z^\d_l,
\ee
and verify that $\sum_i Z_i^\a Z^i_\b = \d^\a_\b,  Z^i_\a Z_j^\a = \d^i_j$
and also $\ov{Z}{}^{i\a} := \Sigma^{\a\b}\bar Z^i_{\b}.$ The condition \Ref{defZ} guarantees that
the twistors $Z_i$ form a basis of $\C^4$. We require it to be orthogonal with respect to the inner product $\Sigma$, i. e.
\be\label{I}
{\cal I}: \quad \r_{ij} \equiv \Sigma(Z_i, Z_j)=0\quad \forall i\neq j,
\ee
so that  $\r_{ij} =2s_i\d_{ij}$ is a diagonal matrix. With (\ref{I})
holding we have 
$ |\ZZ|^2  \stackrel{\cal I}{=} 16s_1s_2s_3s_4$ and $\ov{Z}{}^{i\a}\stackrel{\cal I}{=}  1/({2s_i}) Z_i^\a$,
and various resolutions of the identity:
\be\label{ZIdRes}
\sum_i \f1{2s_i} Z^\a_i \ov{Z}_{i\b} \stackrel{\cal I}{=} \d^\a_\b, \quad
\ov{Z}{}^{i\a} Z^j_\a \stackrel{\cal I}{=} \f1{2s_i} \d^{ij}, \qquad \sum_i 2s_i  Z^i_\a \ov{Z}{}^{i\b} \stackrel{\cal I}{=} \d^\b_\a.
\ee

Consider a second set of four twistors 
$(\tilde{Z}_1, \tilde{Z}_2, \tilde{Z}_3, \tilde{Z}_4)\in
{\mathcal{T}}\equiv \T\times \T\times \T\times \T$.
We assume the twistors within each set to be linearly independent and incident 
- a condition that we still refer to as $\cal I$.
Thus  we have two orthogonal bases for $\C^4$, and we can
construct a matrix that maps one orthogonal basis into the other, 
which will give a dyadic representation of a unitary transformation. 
Consider the GL$(4,\C)$ matrix
\begin{align}\label{Gdef}
& G^\a{}_\b = \sum_i \f{\tl Z^\a_i \ov{Z}_{i\b} }{ \sqrt{2s_i} \sqrt{2{\tl s}_i} }, 
\end{align}
as well as its Hermitian conjugate $G^{\dagger}$ defined by
$\Sigma(G(A), B)=\Sigma(A, G^{\dagger} B)$.
If we further impose the matching of the helicities,
\be\label{h}
h : \quad s_i=\tl s_i,
\ee 
then $G$ maps ${\tilde{Z}}_i$ to ${Z}_i$ and is unitary
on the constraint surface $\hat{\cal C}={\cal I}\cup h$.
To further restrict $G\in \SU(2,2)$ we need the additional constraint 
\[
\Phi := \arg \ZZ - \arg \tl \ZZ =0 
\]
which imposes $\mbox{det}(G)=1$, since $\det G=\ZZ\tilde{\ZZ}/(16\sqrt{s_1s_2s_3s_4\tilde{s}_1 \tilde{s}_2\tilde{s}_3\tilde{s}_4})$. 

Summarizing, the matrix \Ref{Gdef} is unitary with respect to $\Sigma$ when the twistors satisfy the incidences $\cal I$ and helicity matching $h$ conditions, and special unitary when they further satisfy the $\Phi$ condition. These are a total of $4+12+12+1=29$ real conditions on a space of 64 real dimensions, therefore the unitary matrices so described are completely arbitrary.
\subsection{Symplectic structure on $T^*\SU(2,2)$}
\label{sec_ss}
Before presenting our main result, let us fix some notations and provide explicit expressions for the symplectic manifold 
$T^*\SU(2,2)\simeq \SU(2,2)\times \su(2,2)^*$. 
Let $M^{ab}=-M^{ba}$ form a basis of the Lie algebra $\su(4, 2)$
\be
\label{lie22}
[M^{ab}, M^{cd}] = \eta^{ac} M^{bd} -\eta^{ad} M^{bc} + \eta^{bd} M^{ac} - \eta^{bc} M^{ad} =: - f^{abcd}{}_{ef} M^{ef},
\ee
with $a=0, \dots, 5$, and $\eta^{ab}={\rm diag}(-++++-)$.
We parametrize the base manifold with a $\Sigma$--unitary unimodular 
$4 \times 4$ matrix $G^\a{}_\b$, and the algebra with the generators in the fundamental irrep, which can be written as a traceless $4\times4$ matrix themselves using $M^\a{}_\b = \sum_{a<b} M^{ab} \, \G^{ab\a}{}_\b$, where
$\G^{ab}$ are generators of $\mathfrak{spin}(4, 2)$ introduced earlier. 
There are two versions of the isomorphism, taking $M$ to be either left-invariant or right-invariant vectors fields. Choosing the first option for $M$, we denote $\tl M$ the right-invariant vector fields obtained by  adjoint action, 
\be\label{Ad}
\tl M^\a{}_\b = - (GMG^{-1})^\a{}_\b.
\ee
The cotangent bundle carries a natural symplectic structure, with potential given by the inner product between the left- or right-invariant Maurer--Cartan form and the corresponding vector fields (see e.g. \cite{Alekseev94}),
\be\label{OmG}
\Th_{T^*\SU(2,2)} = \tr(\tl M dG G^{-1}) + {\rm c.c.} = \f12 \tr(\tl M dG G^{-1}) - \f12 \tr(M G^{-1} d G) + {\rm c.c.}
\ee 
This results in the following Poisson brackets, 
\begin{subequations}\label{TSU22}\begin{align}
& \{G^\a{}_\b, G^\g{}_\d\} = 0, && \{M^{ab}, G^\a{}_\b\} = i(G\G^{ab})^\a{}_\b, && \{\tl M^{ab}, G^\a{}_\b\} = -i(\G^{ab}G)^\a{}_\b, \\
& \{M^{ab}, \tl M^{cd}\} = 0, && \{M^{ab}, M^{cd}\} = -f^{abcd}{}_{ef} M^{ef}, && \{\tl M^{ab}, \tl M^{cd}\} = -f^{abcd}{}_{ef} \tl M^{ef},
\end{align}\end{subequations}
where $f^{abcd}{}_{ef}$ are the structure constants 
given by (\ref{lie22}).
The brackets in the first line of (\ref{TSU22}) above give the identification of left- and right-invariant vector fields as respectively right and left derivatives.

In the $T^*U(2,2)$ case we have an additional generator, the center of the algebra \Ref{defU},  corresponding to $M^\a{}_\b$ having a trace; and the determinant $\det G$ is a pure phase but not necessarily 1. These two quantities form a canonical pair disentangled from the rest of the algebra \Ref{TSU22}, 
\be\label{Udet}
\{M^{ab},\det G\}=0, \qquad \{U,\det G\}=2i\det G, \qquad \{U,\hat G^\a{}_\b:=\f{G^\a{}_\b}{(\det G)^{1/4}}\}=0.
\ee

\subsection{Symplectic structure on $\T^8$ and reduction to $T^*\SU(2,2)$}

Let us consider $\T^8$, and split the 8 twistors into two sets $Z_i^\a$ and $\tl Z_i^\a$, $i=1,\ldots 4$ with Poisson brackets
\be\label{ZZPB}
\{Z^\a_k, \ov{Z}_{j\b} \} = i \d_{kj}  \d^\a_\b, \qquad \{\tl Z^\a_k, \ov{\tl Z}_{j\b} \} = - i \d_{kj} \d^\a_\b.
\ee
Under these brackets, the scalar products in each set form a closed $\mathfrak{gl}(4,\C)$ algebra\footnote{For the reader familiar with spin foam models, we point out that $\r_{ij}$ are used to construct the holomorphic simplicity constraints introduced in \cite{IoHolo}.},
\begin{subequations}\be\label{rhoalgebra}
\{\r_{mj}, \r_{kl} \} = -i\d_{ml} \r_{kj} + i \d_{jk} \r_{ml}, \qquad \{\tl\r_{mj},\tl \r_{kl} \} = i\d_{ml} \tl\r_{kj} - i \d_{jk} \tl\r_{ml},
\ee
whose centers are $U=\sum_i s_i$ and $\tl U=\sum_i \tl s_i$. 
The other conformal invariant quantities, the holomorphic volumes $\ZZ$ and $\tilde\ZZ$, commute with the off-diagonal scalar products, whereas any helicity shifts the phase:
\[
\nn
\{ \r_{mj}, \ZZ \} = - i \ZZ \d_{mj}, \qquad \{2 s_m, \ZZ \} = -i \ZZ \quad \forall m,
\]
and similarly for the tilded set, but with opposite signs.

We now look for constraints capable of reducing this 64-dimensional symplectic manifold to $T^*\SU(2,2)$.
The unitarity discussion earlier has already identified a candidate set of constraints: the incidence conditions ${\cal I}$, the helicity matching conditions $h$, and the unimodular condition $\Phi$.
The constraint algebra is given by \Ref{rhoalgebra} above together with
\begin{align}
& \{ h_i, h_j\} = 0, && \{ h_m, \r_{jk}\} = - \f i2 \d_{mk}\r_{jm}+\f i2\d_{mj}\r_{mk}, && \{ h_m, \tl\r_{jk}\} = - \f i2 \d_{mk}\tl\r_{jm}+\f i2\d_{mj}\tl\r_{mk}, \\
& \{h_m, \Phi\} = 0, && \{\Phi, \r_{mj}\} = 0 \quad \forall m\neq j, && \{\Phi, \tl\r_{mj}\} = 0 \quad \forall m\neq j.
\end{align}\end{subequations}
These brackets are all zero on the $\cal I$ surface, except for
\begin{align}
& \{\r_{mj}, \r_{jm}\} = 2i(s_m-s_j), && \{\tl\r_{mj},\tl \r_{jm}\} = -2i(\tl s_m-\tl s_j).
\end{align}
Therefore, $h_i$ and $\Phi$ are always first class. The incidences are generically second class;  some or all become first class on measure-zero subsets of the phase space where two or more helicities match. In the generic case, 
symplectic reduction by $h$ and ${\cal I}$ gives a space of dimensions
\[
{\rm dim} (\T^8) - 4\times2 - 12 - 12 = 32= \mbox{dim}(T^*U(2,2)),
\]
and a further reduction by $\Phi$ brings it down to 30$= \mbox{dim}(T^*\SU(2,2))$ .
For the symplectic reduction to work however, we have to remove some regions of the initial phase space. First, our construction of the group element requires non-null twistors, and linearly independent in each sector. Any parallel pair will imply the vanishing of either $\ZZ$ or $\tl \ZZ$ and thus $\det G=0$.
Furthermore, the counting above shows that we want the incidence conditions to be second class, therefore we must exclude within each sector twistors with the same helicity.

Let $\T^8_\star$ be the subspace of $\T^8$ satisfying the following anholonomic restrictions:
\begin{enumerate}[(i)]
\item the twistors within each group of 4 are linearly independent, 
and non--null. 
\item the twistors within each group of 4 have different helicities, 
$s_i\neq s_j$ and  $\tl s_i\neq \tl s_j$ for $i\neq j$.
\end{enumerate}

{\bf Proposition.} \emph{The symplectic reduction of $\T^8_\star$ by the helicity matching and incidence constraints}
\begin{subequations}\begin{align}
& h_i = s_i-\tl s_i = 0, && {\rm (4 \ real, \ first \ class)}\\
& \r_{ij} = 0 = \tl\r_{ij} \quad \forall i\neq j, && {\rm (12 \ complex, \ second \ class)} 
\end{align}\end{subequations}
\emph{describes a symplectic space of 32 real dimensions isomorphic to $T^*U(2,2)$, parametrized by }
\begin{align}\label{GM}
& G^\a{}_\b := \sum_{i=1}^4 \f{\tl Z^\a_i \ov{Z}_{i\b} }{ \sqrt{2s_i} \sqrt{2{\tl s}_i} }, 
&& M^{ab} = \sum_{i=1}^4 \ov{Z}_{i\a} \G^{ab\a}{}_\b Z^\b_i,
&& \tl M^{ab} = -\sum_{i=1}^4 \ov{\tl Z}{}_{i\a} \G^{ab\a}{}_\b {\tl Z}^\b_i, \\
& && U= \sum_{i=1}^4 s_i, && \tl U = -\sum_{i=1}^4 \tl s_i,\nonumber
\end{align}
\emph{with $G^\a{}_\b\in \U(2,2)$, and $(M^{ab},U)$ and $(\tl M^{ab},\tl U)$ respectively left-invariant and right-invariant vector fields isomorphic to the $\u(2,2)$ algebra.}
\smallskip\\

\emph{A Further reduction by the additional constraint }
\be
\Phi = \arg \ZZ - \arg \tl \ZZ =0
\ee
\emph{imposes $\det G=1$, removes $U$ from the phase space, and describes a symplectic space of 30 real dimensions isomorphic to $T^*\SU(2,2)$, parametrized by \Ref{GM} above.
Therefore, 
$
\T_\star^8/\!/{\cal C} \simeq T^*\SU(2,2)
$
with ${\cal C}=h\cup{\cal I}\cup\Phi.$}
\bigskip\\
{\bf Proof.}
To prove the symplectic reduction we need to show that the reduced variables commute with all the constraints, are all independent, and generate the Poisson algebra of $T^*\SU(2,2)$. 
Because of the presence of second class constraints, which we have not explicitly solved, the reduced algebra is defined a priori through the Dirac bracket
\begin{align}
\{F,G\}_{\rm D} := \{F,G\} - \sum_{i\neq j}& \{F,\r_{ij}\} \{\r_{ij},\bar\r_{ij}\}^{-1} \{\bar\r_{ij},G \} + \{F,\bar\r_{ij}\} \{\bar\r_{ij},\r_{ij}\}^{-1} \{\r_{ij},G \}
\nonumber
 \\\nn
& + \{F,\tl\r_{ij}\} \{\tl\r_{ij},\bar{\tl\r}_{ij}\}^{-1} \{\bar{\tl\r}_{ij},G \} + \{F,\bar{\tl\r}_{ij}\} \{\bar{\tl\r}_{ij},{\tl\r}_{ij}\}^{-1} \{{\tl\r}_{ij},G \}.
\nonumber
\end{align}
The only non-vanishing entries of the Dirac matrix are 
\[
\{\r_{mj},\bar\r_{mj}\} = 2i(s_m-s_j), \qquad \{\tl\r_{mj},\bar{\tl\r}_{mj}\} = -2i(\tl s_m-\tl s_j),
\]
thus the Dirac matrix has zeros everywhere except on $2\times 2$ blocks along the diagonal. 
The inverse is then easy to compute, being given by a matrix with the same structure, and elements given by minus the inverse of the original entries. 

For the algebra generators, we have
\begin{align} 
 & \{h_i, M^\a{}_\b \} =  0, &&  \{\r_{ij}, M^\a{}_\b \} = 0, &&  \{\Phi, M^\a{}_\b \} = 0, \nonumber\\
& \{h_i, U \} =  0, && \{\r_{ij}, U \} = 0, && \{\Phi, U \} = 2.\nonumber
\end{align}
The commutation with the second class constraints means that the Dirac bracket for the algebra generators coincides with the Poisson bracket.
For the group elements, we have (with shorthand notation $G^\a_i{}_\b:=  \f{\tl Z^\a_i \ov{Z}_{i\b} }{ \sqrt{2s_i} \sqrt{2{\tl s}_i} }$)
\begin{align}
& \{h_i, G^\a{}_\b \} = 0, \qquad  \{\Phi, G^\a{}_\b \} = \f12\sum_i \left[ \f{\tl Z^\a_i {Z^i_\b}}{\sqrt{2s_i}\sqrt{2\tl s_i}} +
\f{\ov{\tl Z}{}^{i\a} \ov{Z}_{i\b}}{\sqrt{2s_i}\sqrt{2\tl s_i}} 
-G_i^\a{}_\b \left(\f1{2s_i}+\f1{2\tl s_i}\right)\right] \stackrel{{\cal I},h}{=}0,\\
& \{\r_{kj}, G^\a{}_\b \} = i \f{\tl Z^\a_j \ov{Z}_{k\b}}{\sqrt{2s_k}\sqrt{2\tl s_k}} + i \r_{kj} \left(\f{G^\a_k{}_\b}{2s_i} - \f{G^\a_j{}_\b}{2s_j}\right)
\stackrel{{\cal I},h}{=} i \f{\tl Z^\a_j \ov{Z}_{k\b}}{ 2s_k },\label{rG} \\
& \{\tl\r_{kj}, G^\a{}_\b \} = i \f{\tl Z^\a_j \ov{Z}_{k\b}}{\sqrt{2s_k}
\sqrt{2\tl s_k}} + i \r_{kj} \left(\f{G^\a_k{}_\b}{2s_k} - \f{G^\a_j{}_\b}{2s_j}\right)
\stackrel{{\cal I},h}{=} i \f{\tl Z^\a_j \ov{Z}_{k\b}}{ 2s_k } \label{rGtl}.
\end{align}
Even though the group element does not commute with the incidence constraints, 
its Dirac bracket with itself coincides with the Poisson bracket, thanks to opposite contributions from the two sets,
\[
\sum_{i\neq j} \{G^\a{}_\b,\r_{ij}\} \{\r_{ij},\bar\r_{ij}\}^{-1} \{\bar\r_{ij},G^\g{}_\d\}
+ \{G^\a{}_\b,\tl\r_{ij}\} \{\tl\r_{ij},\bar{\tl\r}_{ij}\}^{-1} \{\bar{\tl\r}_{ij},G^\g{}_\d\} \stackrel{{\cal I},h}{=}0.
\]
Therefore, the Dirac bracket of all reduced variables coincides with the Poisson bracket.
Furthermore, this shows also that $G$ and $M$ are are gauge-invariant with respect to all first class constraints in $\T^8_\star$.
We are left to check that they satisfy the right algebra, namely \Ref{TSU22}. 

This means that $(G^\a{}_\b,M^\a{}_\b)$ span the 32 dimensional 
reduced phase space 
We have also already proved that $G$ is unitary,
and we now show that on-shell of the constraints it relates $M$ and $\tl M$ via the adjoint action, since
\[
\tl M^{ab}  \stackrel{{\cal I},h}{=} - \sum_i \ov{Z}_{i\a} \Big(G^{-1} \G^{ab} G\Big)^\a{}_\b Z^\b_i = -(GMG^{-1})^\a{}_\b.
\]
It remains to show that they satisfy the right brackets. To that end, we compute
\begin{align}\label{GG}
\{G^\a{}_\b, G^\g{}_\d \} &= \sum_{kj} \f{\tl Z^\a_k \tl Z^\g_j }{ \sqrt{\tl s_k} \sqrt{{\tl s}_j} } \left\{ \f{\ov{Z}_{k\b}}{ \sqrt{s_k}},  \f{\ov{Z}_{j\d}}{ \sqrt{s_j}} \right\}
+ \f{\ov{Z}_{k\b} \ov{Z}_{j\d}}{ \sqrt{s_k} \sqrt{{s}_j} } \left\{ \f{\tl Z^\a_k  }{ \sqrt{\tl s_k}},  \f{ \tl Z^\g_j }{ \sqrt{\tl s_j}} \right\} \\
&= \f i2 \sum_k \f{\tl Z^\a_k \ov{Z}_{k\b} \tl Z^\g_j \ov{Z}_{j\d}}{ {s_k} {s}_j } \left(s_k^{-1}-s_k^{-1}+\tl s_k^{-1}-\tl s_k^{-1} \right) \equiv 0,\nonumber
\end{align}

\be\label{AppMG}
\{M^{ab}, G^\a{}_\b\} = i(G\G^{ab})^\a{}_\b, \qquad \{\tl M^{ab}, G^\a{}_\b\} = - i(\G^{ab}G)^\a{}_\b.
\ee
As for the brackets of the algebra generators $M$, they follows immediately by linearity from the ones with a single twistor. 
We remark that no constraints were used :the Poisson brackets reproduce the right algebra 
on the whole of $\T^8_\star$. The role of the constraints is truly to restrict the matrix to be unitary and special unitary.

For the final step leading to $T^*\SU(2,2)$, note the Poisson algebra 
we obtained is separable, since $(U, \det G)$ form a canonical pair with brackets \Ref{Udet},
as can be easily verified using 
\[
\{\ZZ , M^{ab}\} = i \ZZ \sum_k Z^k_\a \G^{ab\a}{}_\b Z^\b_k = i \ZZ \, \tr (\G^{ab}) \equiv 0
\]
(except of course if $M=U$ is the $U(1)$ generator, in which case we get correctly $2i\ZZ$).
Then, to reduce to $T^*\SU(2,2)$, we simply impose $\det G=1$ as a (first class, real) constraint, which modules out $U=\sum_i s_i$ as a gauge orbit. Since we already know that two helicities need to have opposite signs, we can fix $U=0$ without loss of generality.
\koniec

Three remarks are in order. Firstly,
when two or more twistors in the same set have the same
helicity, some or all of the incidence constraints become first class. The symplectic reduction describes a smaller phase space not parametrized
by a unitary group element, because
$
\{\r_{ij}, G^\a{}_\b\}\not\approx 0.
$
Secondly, since the helicities can always be made to match in projective twistor    
space, this shows the importance of using the full twistor space for our
symplectic reduction to work.

Finally, instead of working with eight copies of twistor space, we could have picked a pair of self--dual tetrahedra
${\bf T}$ and $\widetilde{\bf T}$ in $\CP^3$ from \S\ref{sd_section}. By construction of these tetrahedra, the incidence constraints
$\rho_{ij}=\tilde{\rho}_{ij}=0$ have already been imposed.  To impose the helicity constraints $h_i=0$ we assign four different colors to
vertices of each tetrahedron, and define $G$ as a $\Sigma$--unitary matrix acting on a configurantion space
of self--dual tetrahedra, and preserving colors of vertices. If we interpret $\ZZ=\ZZ({\bf T})$ as a {\em holomorphic volume}
of the tetrahedron ${\bf T}$, then the final constraint $\Phi=0$ is that $G$ preserves the phase of this holomorphic volume,
which can also  be put in a form
\be
\ZZ = \f1{6} \sum_{i,j,k,l} \eps^{ijkl}I_{\alpha\beta}I_{\gamma\delta}Z^{\alpha}_iZ^{\beta}_jZ^{\gamma}_kZ^{\delta}_l
D_{ijkl},
\ee
where $I_{\alpha\beta}$ is the infinity twistor, and $D_{ijkl}\equiv |X_{ij}-X_{kl}|^2$ are squared distances between the vertices of the 
octahedron from Figure 4 in \S \ref{sd_section} taken with respect to the holomorphic metric on $\M_\C$.

The only other context where a volume of a polygon in the twistor space plays a role in physics is 
the amplituhedron of \cite{nima}. It remains to be seen whether there is any connection between the amplituhedron and our work.

\section{Breaking the conformal symmetry}
\label{conf_sym_sec}
In twistor theory it is common to break the conformal symmetry introducing an infinity twistor, which specifies the asymptotic structure of the conformally flat metric \cite{PenroseRindler2}. The choice of infinity twistor determines if the remaining symmetry is Poincar\'e or the 
(anti-)De Sitter.
Here we are interested instead in a different reduction that takes us directly to $\SL(2,\C)$, since this is the local gauge group of general relativity.
As shown in \cite{IoMiklos}, this reduction can be achieved without using the infinity twistor, but rather requiring conservation of the dilatations between the two sets of twistors.
This means that the we preserve not only the pseudo-Hermitian structure 
$\Sigma$, but also $\g_5$. Since $\g_5$ is the equivalent in the Clifford algebra of the Hodge dual, it is clear that preserving this structure fixes scales. 
And from the $\su(2, 2)$ algebra we see that this condition breaks translations and conformal boosts, allowing only the Lorentz subalgebra.

On the dilatation constraint surface, the description of the remaining Lorentz algebra in $\T^8_\star$ becomes largely redundant. Building on the results of \cite{IoWolfgang}, we know it is enough to work with a pair of twistors only. To eliminate the redundancy, we thus impose the additional constraints
$Z_1=Z_2=Z_3=Z_4$ and $\tilde{Z}_1=\tilde{Z}_2=\tilde{Z}_3=\tilde{Z}_4$.

On-shell of these constraints, $\cal I$ and $\Phi$ become trivial, and $h$ reduces to a single equation. This, together with dilatation constraint 
forms a pair of first class constraints, and we recover  the symplectic reduction
to $T^*\SL(2,\C)$
already established in \cite{WielandTwistors, IoWolfgang, IoTwistorNet}.
The final reduction to $T^*\SU(2)$ relevant to LQG is done introducing a time-like direction, which identifies an 
$SU(2)$ subgroup of $SL(2,\C)$ and a Hermitian structure $||\cdot||^2$. From the twistorial viewpoint, the constraint achieving this reduction is the incidence of two twistors on the same chosen time-like direction. See \cite{IoMiklos} for a review.\footnote{See also \cite{null} for related reductions to the little groups $ISO(2)$ and $SU(1,1)$ stabilizing resp. a null and a space-like direction.}
As a side comment of mathematical interest, it is known \cite{guillemin2006variations} that $T^*SU(2)\cong \C^4/\C^*$ obtained in this way is the maximal co-adjoint orbit of $SU(2, 2)$, and that $SU(2, 2)$ and $U(1)$ form a Howe pair. It may be interesting to establish a precise relation between the Howe pairs,  and the reduction presented in \S \ref{main_sec}.

Coming back to our physical motivations, the work presented has two applications. First,
 the twistorial parametrization of $T^*SU(2,2)$ obtained provides a convenient starting point to construct $SU(2,2)$ spin networks and their holonomy-flux algebra through a generalized Schwinger representation. The flux operators will be the standard holomorphic algebra operators used in quantum twistor theory, whereas the holonomy operators can be built from a suitable operator ordering of \Ref{Gdef}.
Secondly, our classical results are sufficient to deduce how the geometric interpretation of LQG spin networks should be seen from the perspective of $SU(2,2)$ spin networks. The reduction discussed above from $T^*SU(2,2)$ to $T^*SU(2)$ acts trivially on the algebra generators, hence the spin label $j$ describing LQG's quantum of area is simply the $SU(2)$ Casimir with respect to the canonical time-like direction $N^I=(1,0,0,0)$, namely with respect to the canonical 3-vector $(1,0,0,0,1,1)$ in $E^{4,2}$. The effect on the holonomy is less trivial. In particular, the $SL(2,\C)$ matrix element is given by
\be\label{hinG}
h^A{}_B = \f{\tl\om^A \pi_B - \tl\pi^A\om_B}{\sqrt{\pi\om}\sqrt{\tl\pi\tl\om}} \approx
 \f i2 \f{\sqrt{s}}{\sqrt{\pi\om}} \f{\sqrt{\tl s}}{\sqrt{\tl\pi\tl\om}} \Big(G^A{}_B + \eps^{AC} \overline{(G_{\dot C}{}^{\dot D})}\eps_{DB}\Big)
\ee
on-shell of the constraints, where $G^A{}_B$ and $G_{\dot C}{}^{\dot D}$ are the $2\times 2$ diagonal blocks of \Ref{Gdef}.
Here $(\om^A,\pi^A)$ are the spinor constituents of $Z^\a$, and $\pi\om:=\pi_A\om^A$.
The LQG $SU(2)$ holonomy carrying the extrinsic curvature of the quantum space can be recovered from the Lorentz holonomy as explained in \cite{IoWolfgang}, and the embedding \Ref{hinG} shows how it determines the argument of an $SU(2,2)$ spin network. 
From these considerations we can also remark that the LQG area is \emph{invariant} under the $SU(2,2)$ dilatations, whereas the extrinsic geometry is affected, in agreement with \cite{IoMiklos}.

A suggestion in line with Penrose's original program is to introduce a notion of scale not from the Casimirs, but directly from the eigenvalues of the dilatation generator $D$. Such interpretation is at odds with LQG, and we have clarified why. On the other hand, it may be relevant to allow one to extend the spin network construction of the Hilbert space of loop quantum gravity to more general theories like Poincare gauge theory of gravity or conformal gravity.

\subsection*{Acknowledgements} The work of MD has been partially supported by STFC
consolidated grant no. ST/P000681/1.  ML acknowledges grant no. 266101 by the Academy of Finland. SiS is grateful to Sergey Alexandrov and Kristina Giesel for discussions on symplectic reductions and Dirac brackets. MD thanks Dmitry Alekseevsky
and Nick Woodhouse for useful correspondence.


\end{document}